\documentclass{article}
\usepackage{emulateapj,onecolfloat,graphics}

\lefthead{Vale \& White}
\righthead{Simulating weak lensing}

\begin{document}
\twocolumn[
\title{Simulating weak lensing by large scale structure}
\author{Chris Vale and Martin White}
\affil{Departments of Astronomy and Physics, University of California,
Berkeley, CA 94720}

\begin{abstract}
\noindent
We model weak gravitational lensing of light by large-scale structure
using ray tracing through N-body simulations.
The method is described with particular attention paid to numerical
convergence.  We investigate some of the key approximations in the
multi-plane ray tracing algorithm.  Our simulated shear and convergence
maps are used to explore how well standard assumptions about weak lensing
hold, especially near large peaks in the lensing signal.
\end{abstract}

]

\rightskip=0pt
\section{Introduction}

Weak gravitational lensing is becoming an indispensable tool in modern
cosmology.
Building on early work by Gunn~(\cite{hhm10}), Blandford et al.~(\cite{hhm6}) 
and Miralda-Escude~(\cite{hhm23}) showed that the shear and magnification
are on the order of a few percent in popular cosmologies, which is now
within reach of observations.  Indeed detections of shear correlations
by several groups (Bacon et al.~\cite{hhm1}; Van Waerbeke et al.~\cite{hhm34};
Rhodes et al.~\cite{hhm27}; Hoekstra et al.~\cite{hhm15}; Brown et 
al.~\cite{hhm7}; Jarvis et al.~\cite{hhm20}) indicate the dramatic progress 
being made in observational power and precision.  Other projects, such as 
the CFHT legacy survey\footnote{http://www.cfht.hawaii.edu/Science/CFHLS},
the Deep Lens Survey\footnote{http://dls.bell-labs.com/},
and NOAO deep survey\footnote{http://www.noao.edu/noao/noaodeep/}
and proposed projects such as SNAP\footnote{http://snap.lbl.gov},
DMT/LSST\footnote{http://www.dmtelescope.org/dark\_home.html},
and Pan$-$STARRS\footnote{http://www.ifa.hawaii.edu/pan-starrs/}
will continue this trend as they map the shear on large fractions of the sky.
In order for the full scientific return from lensing to be realized, both
fundamental theory and data analysis techniques must keep pace with these
observational advances.

In principle the effect exploited by these surveys is simple: gravitational
lensing of light rays by foreground large-scale structure magnifies and shears
the images of background galaxies, and this distortion can be mapped to
provide information on the matter distribution and cosmological model.
Like the cosmic microwave background (CMB), the theory of weak lensing is
simple and clean
(see Bartelmann \& Schneider~\cite{hhm4} or Mellier~\cite{hhm22} for a review).
Unlike the the CMB, however, lensing on sub-degree scales probes large-scale
structure in the non-linear regime and a full description has not been
achieved through purely analytic means.
Fortunately, on scales relevant to weak lensing, the hierarchical growth of
structure can be accurately simulated using N-body methods.
The lensing effect of this structure can then be computed directly by ray
tracing through the simulation.
In this paper we describe an implementation of such a method, building upon
the work of
(Blandford et al.~\cite{hhm6}; Wambsganss, Cen \& Ostriker~\cite{hhm36};
Couchman, Barber, \& Thomas~\cite{hhm8}; Fluke, Webster, \& 
Mortlock~\cite{hhm9}; Hamana, Martel \& Futamase~\cite{hhm11}; Jain, Seljak, 
\& White~\cite{hhm19}; White \& Hu~\cite{hhm38}; Barber, Thomas, Couchman, 
\& Fluke~\cite{hhm2}; Hamana \& Mellier~\cite{hhm12}; Hennawi et 
al.~\cite{hhm14}; Padmanabhan, Seljak, \& Pen~\cite{hhm25}).
Our goal is to quantify the numerical precision of such algorithms, and to
investigate the computational cost of achieving a given theoretical precision.

The outline of the paper is as follows.
In \S\ref{sec:formalism} we introduce the weak lensing formalism and in
\S\ref{sec:raytracing} we describe our implementation of a multi-plane
ray-tracing algorithm.
In \S\ref{sec:results} we present some basic results for comparison to
previous work while in \S\ref{sec:numerical} we evaluate the numerical
convergence of the algorithm.  Some interesting physical results are given
in \S\ref{sec:physical} before we conclude in \S\ref{sec:conclusions}.

\section{Weak lensing formalism} \label{sec:formalism}

We make use of a multiple lens plane algorithm in order to simulate the
distortion and magnification effect of foreground matter on background light
rays.  In the following section, we provide a summary of equations directly
relevant to this paper (see Jain, Seljak, \& White~(\cite{hhm19}) or 
Schneider, Ehlers, \& Falco~(\cite{hhm28}) for a more thorough treatment) 
and then describe our ray tracing method.  Note that we adopt units where 
$c=1$, and we will work in comoving coordinates.

In a universe governed by the Robertson-Walker metric,
the change in direction of a light ray propagating through space is:
%
%
\begin{eqnarray}
d {\vec \alpha} = -2 \nabla_\bot\phi\  d\chi
\label{eqn:einstein}
\end{eqnarray}
\noindent
where $d \vec \alpha$ is
the bend angle, $\nabla_\bot$ denotes the spatial gradient perpendicular
to the path of the light ray, $\phi$ is the 3 dimensional peculiar
gravitational potential, and $\chi$ is the radial comoving coordinate.
The change in position on a plane perpendicular
to the line of sight at a position $\chi$ due to a deflection
$d \vec \alpha$ at $\chi '$ is:
%
%
\begin{eqnarray}
  {d \vec x( \chi ) = r( \chi - \chi ') d \vec \alpha (\chi ')}
\label{eqn:dx}
\end{eqnarray}
where $r(\chi)$ is the comoving angular diameter distance.
Integrating to get $\vec x(\chi)$ and then dividing by $r(\chi)$ yields:
%
%
\begin{equation}
  {\vec \theta(\chi)}  = {-2 \over { r(\chi)}}
     {\int_0^{\chi}d\chi ' \ r( \chi - \chi ') \nabla_\bot \phi +
     \vec \theta (0)}
\label{eqn:theta}
\end{equation}
\noindent
where $\vec \theta(\chi)$ is the angular position of the light ray at
comoving coordinate $\chi$.  The $2\times 2$ distortion
matrix, ${A_{ij} \equiv} {\partial \theta_i (\chi) / \partial \theta_j (0)}$,
is given by:
%
%
\begin{eqnarray}
  A_{ij} = {-2} {\int_0^{\chi}{ d \chi ' \ g(\chi ', \chi)
           \nabla_i \nabla_j \phi}} + \delta_{ij}
\label{eqn:Aij}
\end{eqnarray}
\noindent
where we make use of the definition:
%
%
\begin{eqnarray}
  g(\chi ', \chi) \equiv {r(\chi - \chi ') r(\chi ') \over r(\chi)}
\label{eqn:g(chi)}
\end{eqnarray}
This matrix is normally decomposed as:
%
%
\begin{eqnarray}
  {\bf A} = \pmatrix{1-\kappa-\gamma_1&-\gamma_2-\omega\cr
            -\gamma_2+\omega&1-\kappa+\gamma_1}
\label{eqn:Avec}
\end{eqnarray}
\noindent where $\kappa$ is the convergence, $\gamma$ is the shear, and
$\omega$ is the rotation.

The gravitational potential, $\phi '$, is related to the mass density,
$\rho$, through Poisson's equation:
%
%
\begin{equation}
  \nabla^2 \phi ' = 4 \pi G a^2 \rho
\label{eqn:poisson1}
\end{equation}
\noindent
where the gradient is with respect to comoving coordinates (that's why
there's an $a^2$ on the right hand side) and $\phi '$
is the sum of a smooth background potential $\bar \phi$ and the local
peculiar potential $\phi$.  We can subtract the background terms and use
$\nabla^2 \bar \phi = 4 \pi G \bar \rho$, where $\bar \rho$ is the mean
density in the region, and substitute $a^3 \bar \rho = \bar \rho_0$ to find
%
%
\begin{equation}
  \nabla^2 \phi = {4 \pi G \bar \rho_0 \over a}
  \left( {\rho\over \bar\rho}-1 \right) \qquad .
\label{eqn:poisson2}
\end{equation}
If we then define the relative mass overdensity,
$\delta \equiv \rho/\bar\rho-1$, and substitute
$\bar\rho_0 = 3 H_0^2 \Omega_m / 8 \pi G$ we obtain
%
%
\begin{equation}
  \nabla^2 \phi = {{3 H_0^2 \Omega_m} \delta \over 2a} \qquad .
\label{eqn:lapchi}
\end{equation}

In the discrete approximation, the interval between the source and the
observer is divided into distinct regions separated by a comoving
distance $\Delta\chi$.  The relative mass overdensity in a given region
is then projected onto a lens plane within the region and perpendicular
to the line of sight.  An effective two dimensional gravitational potential,
$\psi=2\int \phi\,d\chi$, is defined on the this plane, enabling
a two dimensional Poisson equation relating $\psi$ and the matter
content in the region to be written as
%
%
\begin{equation}
  \nabla^2 \psi = {3 H_0^2 \Omega_m} \Sigma
\label{eqn:lappsi}
\end{equation}
where $\Sigma = \int\delta\,d\chi$ is the projected two
dimensional relative mass overdensity.  Eqs (\ref{eqn:theta}) and
(\ref{eqn:Aij}) then become
%
%
\begin{eqnarray}
{\vec \theta_n} = - {\sum_{m=1}^{n-1} {r( \chi_n - \chi_m) \over
r( \chi_n)} \nabla_{\bot} \psi_m} + \vec \theta_1
\nonumber \\
{\bf A}_n  = {\bf I} - \sum_{m=1}^{n-1} g(\chi_m , \chi_n) {\bf U}_m {\bf A}_m
\label{eqn:discrete}
\end{eqnarray}
%
%
\noindent
where ${\bf U}_m$ is the shear tensor in the $m^{th}$ region defined by
\begin{equation}
U_{ij} \equiv {\partial^2 \psi_m \over \partial x_i \partial x_j}
\label{eqn:Uij}
\end{equation}
It will be useful to decompose the distortion matrix at a given plane into
two components (Seitz, Schneider, and Ehlers~\cite{hhm29})
${\bf B}_n$ and ${\bf C}_n$ such that for a zero curvature universe
(assumed hereafter):
%
%
\begin{eqnarray}
  {\bf A}_{n+1} \equiv {\bf I} - {\bf B}_n + {\bf C}_n \nonumber \\
  {\bf B}_n = {\bf B}_{n-1} + \chi_n {\bf U}_n{\bf A}_n \nonumber \\
  {\bf C}_n =
  {\chi_n \over \chi_{n+1}} ({\bf C}_{n-1} + \chi_n {\bf U}_n{\bf A}_n)
\label{eqn:decompose}
\end{eqnarray}

\section{ray-tracing algorithm} \label{sec:raytracing}

The goal of our weak lensing algorithm is to simulate the deflection and
distortion that light rays would experience as they propagate through an
intervening matter distribution that is statistically similar to that of the
real universe.  In this section we describe two different implementations of
the weak lensing algorithm: first, a standard implementation based on multiple
projected-mass lens planes, and then a second, three-dimensional version.
The modeling of the mass distribution is discussed in \S\ref{sec:numerical}.

\subsection{Standard (2D) algorithm}

The standard implementation consists of creating a two-dimensional projected
mass plane from an N-body simulation, computing the position, propagation
direction, and distortion matrix on the plane for each ray, and then repeating
the process until the rays have been traced from the observer to the source.
Here is a more detailed description:

\noindent
1. We begin by using an N-body simulation (which we describe later) to create
the structure responsible for the lensing.  Since lensing is only sensitive to
density fluctuations perpendicular to the line of sight, the simulation must
resolve structures on length scales $\sim \chi \theta$, where $\chi$ is a
typical survey depth, and $\theta$ is the desired angular resolution.
Light-rays from redshifts of interest must therefore travel thousands of Mpc
through structures resolved below Mpc scales.

It is currently computationally very expensive to simulate both scales
simultaneously, but, fortunately, this is not necessary.  Since there is little
lensing power on very large scales, it makes no sense to simulate huge boxes
whose pieces are almost independent.  We employ a technique (common in the
literature) in which we simulate a volume which is large enough to be a fair
sample of the universe, yet small enough to provide sufficient small scale
resolution at reasonable cost.  In this simulation, mass particles in a box
evolve in time under the influence of gravity, and the positions of the
particles are recorded at time intervals corresponding to regular intervals in
comoving distance.  These output boxes are then used to create the mass
distribution to be traced through.  Note that since each box is actually the
same matter distribution at different stages of evolution, care must be taken
to avoid tracing over the same structures more than once.

\noindent
2. A selected number of light rays to be traced are then initialized at the
observer for a square field of view.  For each ray, we track the position $\vec
x$, the propagation direction $\vec \alpha_n$ relative to the line of sight,
and the matrices $\bf B$ and $\bf C$ from equation~(\ref{eqn:decompose}).  The
recursion relations (Eq.~\ref{eqn:discrete}) for ${\bf A}_n$ and $\vec
\theta_n$ require as inputs the previous values for $\nabla_{\bot} \psi$ and
${\bf U}$, all of which need to be stored.  However, we make use of a less
memory intensive algorithm for the distortion matrix based on
Eq.~(\ref{eqn:decompose}) in which we track the values the decomposed
distortion matrix $\bf B$ and $\bf C$ for each ray.  Each iteration requires as
a new input only the current value of ${\bf U}$ at each ray location, which may
then be discarded.   Similarly, by tracking $\vec \alpha_n$, we require only
one value of $\nabla_{\bot} \psi$ at a time.

\noindent

3. Next, we select the source redshift and the number of lens planes we wish to
use.  Note that the sources are all initialized at the same redshift, and the
planes are equally spaced in comoving distance.  This sets the comoving
distance between planes, $\Delta\chi$.  We create a lens plane from an N-body
simulation box using the following technique. We select the N-body particle
dump closest in comoving distance to that of the desired lens plane, and then
randomly choose the $x$, $y$, or $z$ axis as the propagation axis, $\chi$.  A
two-dimensional $\rm N \times N$ lens plane grid is then created by projecting
mass particles from the N-body simulation in a direction parallel to the $\chi$
axis onto a plane at the desired comoving distance, and then recording the mass
onto the grid locations using a cloud in cell (CIC) assignment scheme.   All of
the particles in the box within ${1 \over 2}\Delta\chi$ of the desired plane
are used to make the grid, which is then normalized by the mean density.  The
projected field is thus a square of side length equal to that of the box.

The origin of the lens plane is randomly selected.  This origin shift is
facilitated by the fact that the N-body simulation uses periodic boundary
conditions, and we can use this periodicity to re-map the mass density about
any origin we choose.  This process of rotation and origin shift makes it
unlikely that the same structures will be traced repeatedly.  Note that both
the origin and orientation of the axes are maintained until the box is traced
all the way through, at which point the box is re-randomized, so that
consecutive lens planes are made from a matter distribution that is continuous
everywhere except at the box boundaries.

A variation of this technique, to our knowledge presented here for the first
time, and which we make use of for comparison, avoids introducing these
discontinuities entirely.  The propagation direction is selected at an angle
relative to one of the box axes.  The rays are initialized about a central line
of sight along this direction, and the lens planes are created perpendicular to
the central line of sight and with their origins on it.  The angle is chosen so
that the rays, traced from the observer to the source, never propagate through
the same volume twice.  However, since the boxes are never rotated, nor is
there any random origin shift, the lens planes are made from a matter
distribution that is continuous everywhere.

\noindent
4. The two component gradient of the potential $\nabla \psi$ and the four
components of the matrix $\bf U$ are then computed on the grid.  This is done
by first taking the Fourier transformation of the mass density plane, then
taking the first or second derivatives, respectively, and computing the inverse
transformation.  We do this by making use of the
relation of real space derivatives in Fourier space, for example
\begin{eqnarray}
{\partial \psi (x) \over \partial x} =
-i \int {k_x \tilde \psi (k) e^{-i \vec k \cdot \vec x} d^2k}
\end{eqnarray}
\noindent
5. The position $\vec x_n$ for each ray on the plane is computed using the
stored values $\vec x_{n-1}$ and $\vec \alpha_{n-1}$.  Values for $\bf U$ and
$\nabla \psi$ are then assigned at each ray position, again using a CIC
assignment scheme, and the new deflection angle $\vec \alpha_n = \vec
\alpha_{n-1} +\Delta\vec{\alpha}_n$ is then computed where
%
%
\begin{eqnarray}
\Delta\vec{\alpha} = -{\nabla \psi}
\label{eqn:dangle}
\end{eqnarray}
\noindent
follows from Eq.~(\ref{eqn:einstein}).  The distortion matrix ${\bf A}_n$ is
temporarily created from stored values of ${\bf B}_{n-1}$ and ${\bf C}_{n-1}$
and used along with ${\bf U}_n$ to create the new values ${\bf B}_n$ and ${\bf
C}_n$ for each ray on the plane.

\noindent
6. The process is repeated until each ray is traced all the way back to its
source, thus ensuring that all the rays will meet at the observer, and the
final angular position and distortion matrix are recorded for each ray.

Clearly, one limitation of any algorithm based on lens planes is that the light
rays themselves do not all propagate in precisely the same direction and cannot
all be perpendicular to a given lens plane.  The weak lensing formalism of the
previous section calls for the calculation of derivatives perpendicular to the
light rays propagation direction, so the multiple lens plane algorithm is
necessarily only approximate in this respect.

\begin{figure*}
\begin{center}
\resizebox{3.6in}{!}{\includegraphics{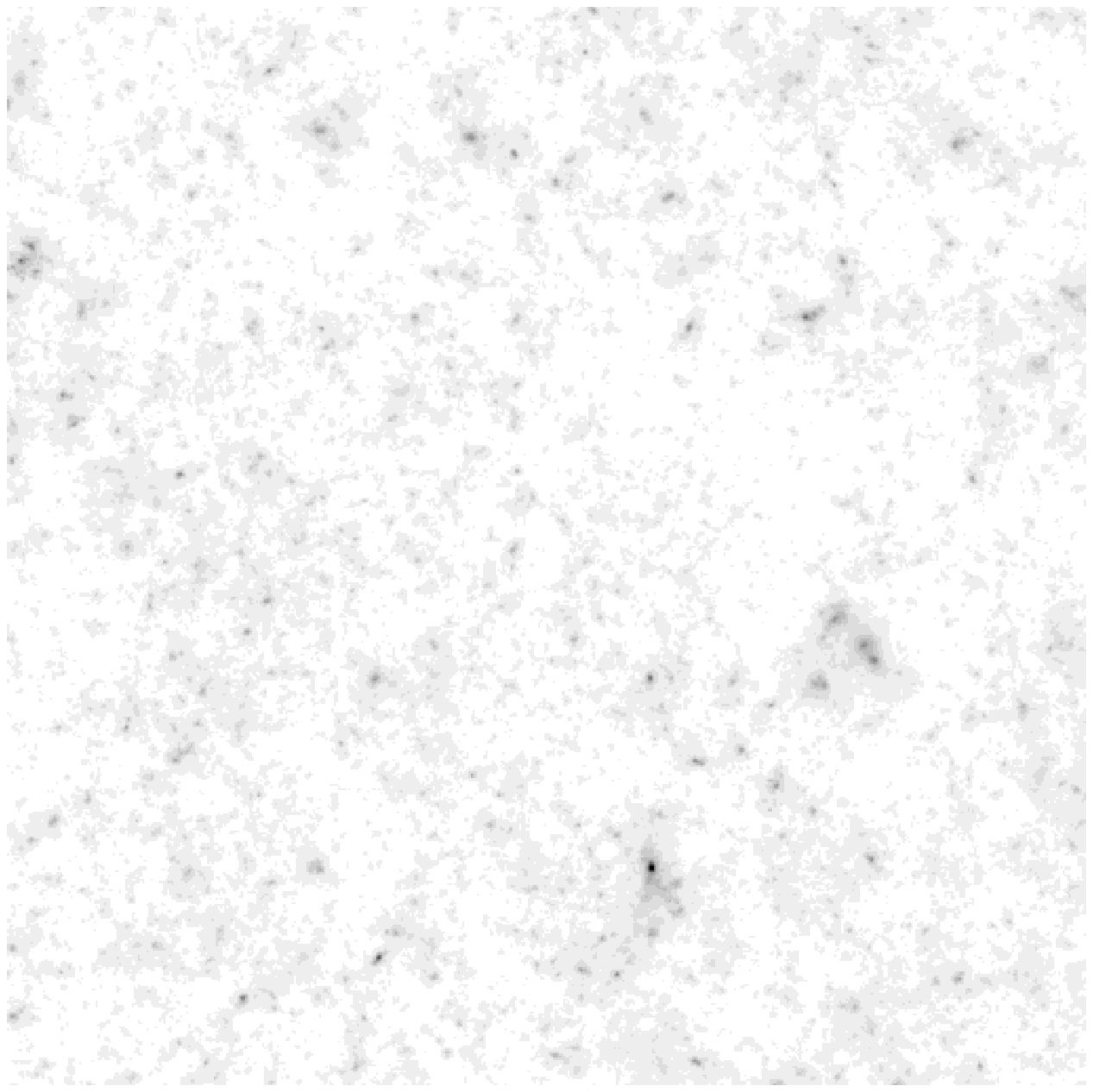}}
\resizebox{3.6in}{3.6in}{\includegraphics{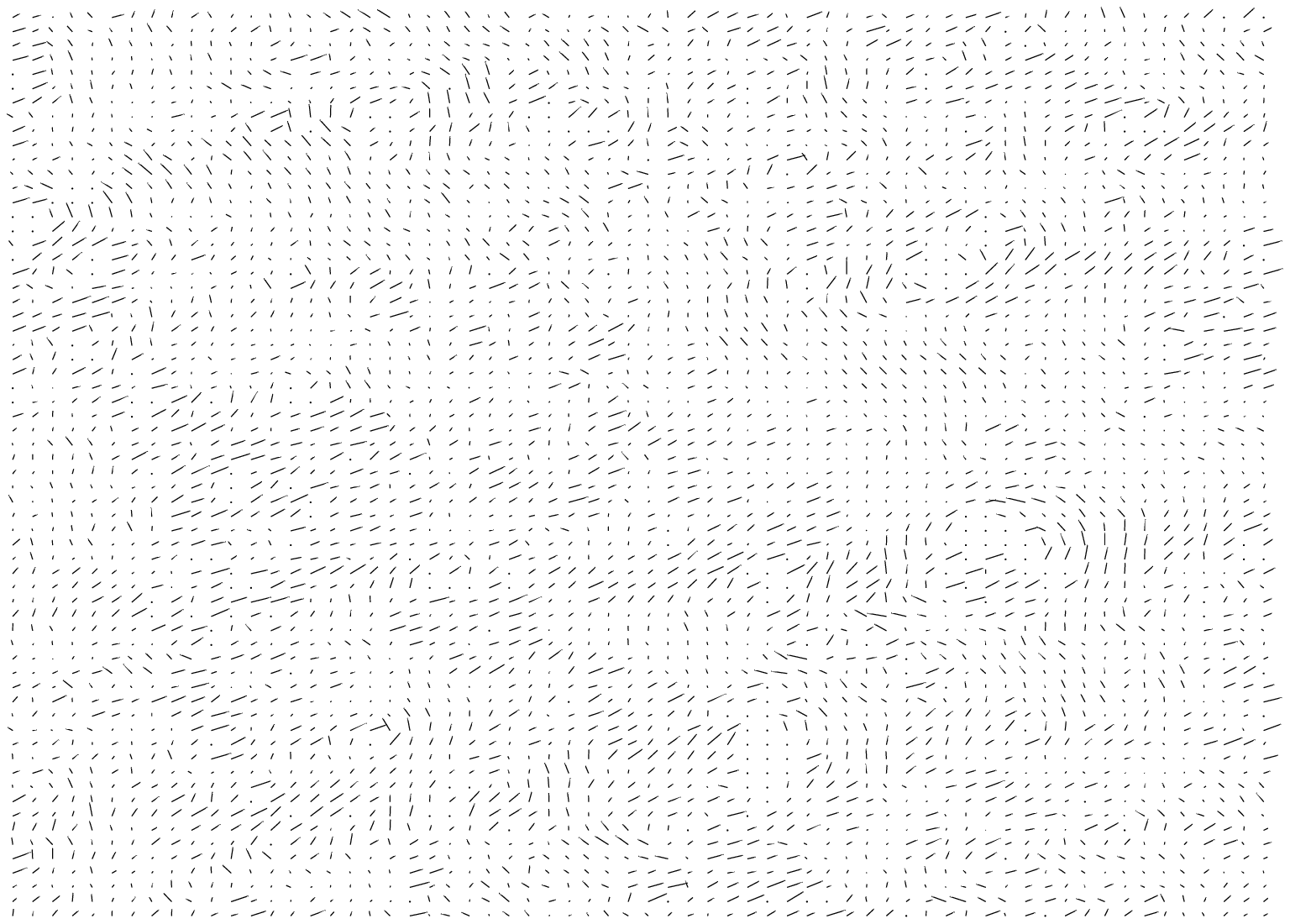}}
\end{center}
\caption{(left) An image of the convergence $\kappa$ from a single realization
for a $2^\circ$ field of view.  The greyscale is linear in $\kappa$, ranging
from white ($\kappa=-0.04$) to black ($\kappa\ge 0.7$).
Regions of high convergence are due to massive structures (typically
clusters of galaxies) along the line-of-sight.
(right) The shear field $\gamma$ for the same realization, exaggerated for
clarity (the magnitude of the shear is at the percent level).  Note that the
shear is tangential around regions of high convergence.}
\label{fig:kmap}
\end{figure*}

\subsection{Extended (3D) algorithm}

We have created a second version of the ray-tracing algorithm that is fully
three-dimensional, and therefore doesn't suffer from this drawback.  It is
quite similar to the standard implementation in most respects, and makes use of
the same N-body particle dumps.  Here is a brief description.

\noindent
1. The source redshift is chosen and light rays in a square field of view are
initialized at the observer.  As before, we will track the position and
propagation direction of each ray, and also the matrices $\bf B$ and $\bf C$.
Note that these are still $2 \times 2$ matrices, since, like the matrix $\bf
A$, they still describe the distortion of a two-dimensional image.

\noindent
2. The matter distribution from the particle dump nearest in redshift to the
current position is mapped onto a three-dimensional regular grid using a CIC
assignment scheme, and the grid is then normalized by the mean density.  As
before, it's important not to re-trace the same structures, so the box
orientations and origins are randomly reassigned, when necessary, to avoid
this.

\noindent
3. The relationship in Fourier space between $\phi$ and $\delta$ is used to
calculate the three first derivatives and six independent second derivatives of
$\phi$ at each point on the grid.  These derivatives are then evaluated using
the CIC method at each ray position.  The components perpendicular to each ray
direction are calculated at each ray position and used to update the values of
$\bf B$, $\bf C$ and $\vec \alpha_n$.   The rays are then projected forward by
a pre-selected comoving distance, and the derivatives are then evaluated at the
new positions.  This process is repeated until it becomes time to use the next
particle dump.

\section{basic results} \label{sec:results}

Before we discuss the numerical convergence of our technique, we present some
basic science results.  These are broadly in agreement with similar studies
presented by (Jain, Seljak, \& White~\cite{hhm19}; Hamana, Martel, \& 
Futamase~\cite{hhm11}; White \& Hu~\cite{hhm38}) as we elucidate later.
Fig.~\ref{fig:kmap} is an example of the convergence ($\kappa$) and shear
($\gamma$) maps that result from the ray-tracing procedure.  The maps are
$2^{\circ}$ on a side and contain $2048^2$ lines of sight.
The average magnitude of the convergence and shear is about $1.8\%$, close to
the level predicted by Blandford et al.~(~\cite{hhm6}) and 
Miralda-Escude~(~\cite{hhm23}).

\begin{figure}
\begin{center}
\resizebox{7cm}{!}{\includegraphics{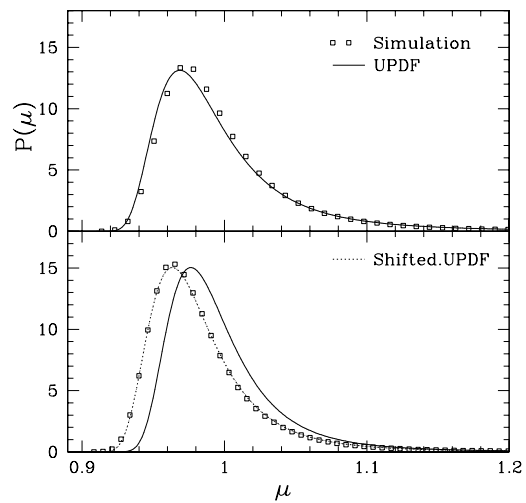}}
\end{center}
\caption{(top) A typical magnification PDF from our simulation (squares)
compared with the prediction (solid line) using the method of
Wang, Holz, \& Munshi~(\cite{hhm37}).   The curves typically agree to within
$\sim 5\%$ near the peaks and $\sim 20\%$ in the high $\mu$ tails.
(bottom) A second example where the model PDF is shifted on the horizontal
axis with respect to that of the simulation.  This occurs frequently in
our maps.  The shape of the measured PDF is still well fit by the prediction
of the model (solid line), as can be seen when this curve is shifted in
$\mu$ (dashed line).}
\label{fig:holzupdf}
\end{figure}

In the following section we describe these fields.  We begin by presenting
the 1-point distribution, and demonstrate that there are not likely to be
any unlensed images at high redshift for this cosmology.
We then present the power spectrum, which we find to be consistent with the
Born and Limber approximations.
We then discuss the skewness ($S_3$) and the kurtosis ($S_4$), which are
given as a function of angular smoothing scale.
We conclude with a description of the convergence peaks in our maps.

The lowest order statistic for the maps is the distribution of the shear
amplitude or convergence.  We show two typical examples in
Fig.~\ref{fig:holzupdf} in terms of the magnification, which in the weak
lensing limit is simply $\mu=1+2\kappa$.  This is compared to the `universal'
form of Wang, Holz, \& Munshi~(\cite{hhm37}) in \S\ref{sec:physical}.
The average magnitude of the convergence over 10 realizations is 0.018,
consistent with the expectation of a weak lensing effect on the order of
a few percent.  The mean and variance of the convergence are -0.002 and
$7.5 \times 10^{-4}$, respectively.
The minimum $\kappa$ for any of our realizations was $-0.049$, which is
greater than the `empty beam' value (an empty beam is a path from the source
to the observer which is sufficiently void of matter that gravitational
interactions may be ignored) of $-0.064$ for our cosmology, consistent with
the fact that all the light rays in our simulations are lensed to some extent.

\begin{figure}
\begin{center}
\resizebox{7cm}{!}{\includegraphics{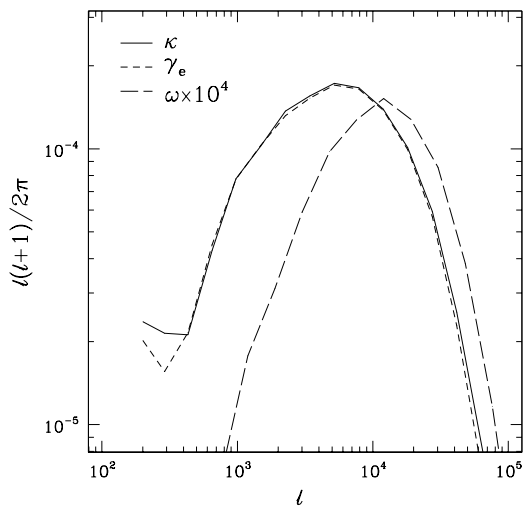}}
\end{center}
\caption{The angular power spectra of the convergence $\kappa$, curl-free
shear $\gamma_E$, and rotation $\omega$ (multiplied by $10^4$) from one of
our simulated maps.  As predicted for the weak lensing regime, the spectrum
of $\omega$ is much less than the spectra of $\kappa$ and $\gamma_E$, which
are nearly equal.}
\label{fig:kgwspectra}
\end{figure}

The shear correlations, from which cosmological inferences are usually drawn,
are shown in Fig.~\ref{fig:kgwspectra} in terms of angular power spectra.
Figure \ref{fig:kgwspectra} shows the power spectrum of ($\kappa$) vs.~the
spectra of the E-mode of the shear ($\gamma_E$) and of the rotation
($\omega$) for one of our simulated maps.  The shear E-mode is defined to
be the curl free component of the shear tensor, and is easily calculated
in Fourier space using
%
%
\begin{equation}
{\tilde \gamma_E} (\vec k) = {{(k_x^2 - k_y^2) \tilde \gamma_1 (\vec k) +
2 k_x k_y \tilde \gamma_2 (\vec k)} \over {k_x^2 + k_y^2}}
\label{eqn:gamma_e}
\end{equation}
where the tilde denotes the Fourier transform.  The `excess' power in
$\kappa$ and $\gamma_E$ at $\ell\sim 200$ is just a fluctuation in this
particular map.

One of the most important predictions of the weak lensing approximation is that
the shear matrix $\bf A$ can be well approximated using the 2nd derivatives of
a scalar potential, as would be the case if only a single lens plane were used
to compute it.  Then $\bf A$ would be symmetric, $\omega$ would be zero, and
$\gamma_E$ would equal $\kappa$.  Thus, in the weak lensing regime, the spectra
of $\kappa$ and $\gamma_E$ are predicted to be nearly equal to each other, and
to be much larger than that of $\omega$.  We agree with earlier work (e.g. JSW)
in verifying this basic assumption of weak lensing (Fig.~\ref{fig:kgwspectra}).

\begin{figure}
\begin{center}
\resizebox{7cm}{!}{\includegraphics{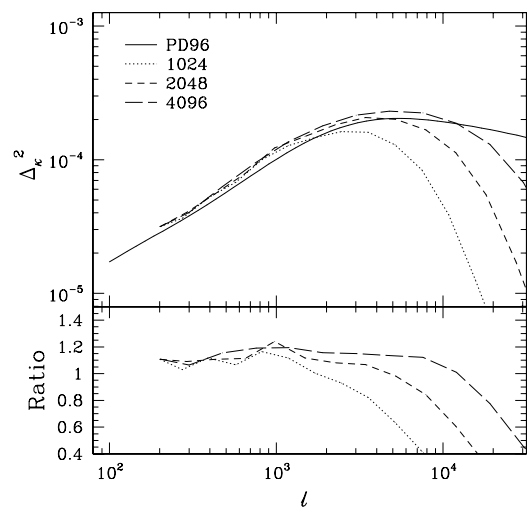}}
\end{center}
\caption{(top) Comparison of the convergence angular power spectra from the
simulations at varying levels of grid resolution with the semi-analytic
prediction computed using the method of Peacock \& Dodds~(\cite{hhm26}) and
Eq.~(\ref{eqn:massspectrum}).  The spectra are averages from 10 realizations of
our high resolution simulation, with $\Delta_{\kappa}^2 \equiv l(l+1)C_l/2\pi$,
for grids of $1024^2$, $2048^2$, and $4096^2$ points, and a side length of
$300\,h^{-1}$Mpc.
(bottom) The ratio of the spectra from our simulations with the semi-analytic
prediction.}
\label{fig:nspectra}
\end{figure}

Figure \ref{fig:nspectra} shows the angular power spectrum of $\kappa$ (at
various levels angular resolution) compared to the semi-analytic result
which makes both the Born and Limber approximations.  This result was first
derived by Kaiser~(\cite{hhm21}) and extended by Jain \& 
Seljak~(\cite{hhm18}).  We present it here in the form of White \& 
Hu~(\cite{hhm38})
%
%
\begin{equation}
\Delta_{\kappa}^2 (l) = {9 \pi \over 4 l} \Omega_m^2 H_0^4
  \int\,\chi'\,d\chi'\ 
  \left[{g(\chi',\chi)\over a(\chi')}\right]^2 \Delta_{\rm m}^2(k,a)
\label{eqn:massspectrum}
\end{equation}
where $g(\chi',\chi)$ is defined in Eq.~(\ref{eqn:g(chi)}),
$\Delta_{\rm m}^2(k) = k^3 P(k)/(2 \pi ^2)$ is the contribution to the
mass variance per logarithmic interval in wavenumber and
$\Delta_{\kappa}^2 (\ell) = \ell^2 C_\ell /(2 \pi)$ is the contribution to
$\kappa_{\rm rms}^2$ per logarithmic interval in angular wavenumber $\ell$.

\begin{figure}
\begin{center}
\resizebox{7cm}{!}{\includegraphics{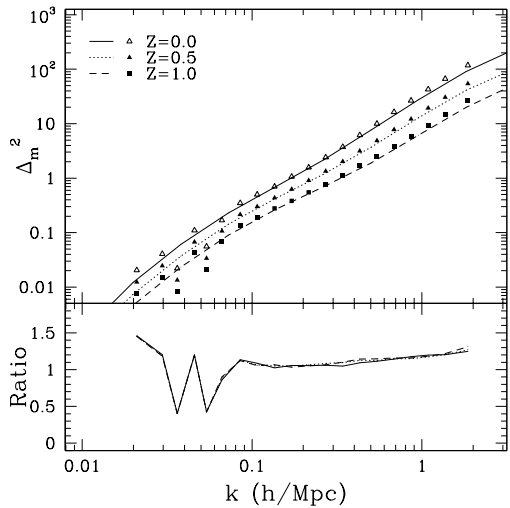}}
\end{center}
\caption{(top) The 3-dimensional matter power spectrum of 3 boxes at selected
redshifts for our high resolution simulation, compared against the
semi-analytic prediction using the prescription of Peacock \& 
Dodds~(\cite{hhm26}).  (bottom) The ratios of the simulated spectra with 
the semi-analytic prediction.}
\label{fig:3dspectra}
\end{figure}

The agreement between the numerical and semi-analytic predictions is quite
good, but not perfect.  One reason for this could be that in evaluating
Eq.~(\ref{eqn:massspectrum}) we used the method of Peacock \& 
Dodds~(\cite{hhm26}) to compute the non-linear power spectrum $P(k)$.
As can be seen in Figure~\ref{fig:3dspectra}, where we compare this
prediction with the measured 3-dimensional matter power spectrum of
our N-body simulation, the Peacock \& Dodds~\cite{hhm26}) fitting formula 
is not exact.  If we replace the $\Delta_{\rm m}^2(k)$ in 
Eq.~(\ref{eqn:massspectrum})
with the values actually measured in our simulation, the discrepancy
disappears.  We conclude therefore that the Born and Limber approximations
are consistent with the power spectrum results from our simulations
(we shall return to this issue in \S\ref{sec:physical}).

\begin{table}
\begin{center}
\begin{tabular}{cccccccc}
   &   &$\sigma_8$ &$S_3(1')$ &$S_3(4')$ &$S_4(1')$ &$S_4(4')$ \\ \hline
JSW   &\vline  &  0.9      &   145    & 140    & 3.8    & 3.1      \\
HM    &\vline  &  0.9      &   118    & 110    & 2.9    & 2.6      \\
WH    &\vline  &  1.2      &    -     & 110    &  -     & 2.8      \\
H2    &\vline  &  0.9      &   138    & 114    & 4.5    & 3.4      \\
VWa   &\vline  &  1.0      &   138    & 120    & 4.1    & 2.8      \\
VWb   &\vline  &  0.8      &   155    & 140    & 5.8    & 4.6      \\ \hline
v.W   &\vline  &  0.9      &   140    & 127    &  -     &  -       \\
TJ    &\vline  &  0.9      &   137    & 147    & 3.5    & 4.2
\end{tabular}
\end{center}
\caption{The skewness and kurtosis reported by several groups using
simulations and analytical predictions (see text).  All are for
$\Omega_m = 0.3 = 1-\Omega_\Lambda$ and $h = 0.7$, and with $S_4$ given
in units of $10^4$.  The values for v.W are for a semi-analytic
interpolation between the predictions of Hyper-Extended Perturbation Theory
(HEPT) and the perturbation theory result, while TJ report values using
the Halo model approach.  All other values are for simulations.}
\label{tab:s3s4}
\end{table}

\begin{figure}
\begin{center}
\resizebox{7cm}{!}{\includegraphics{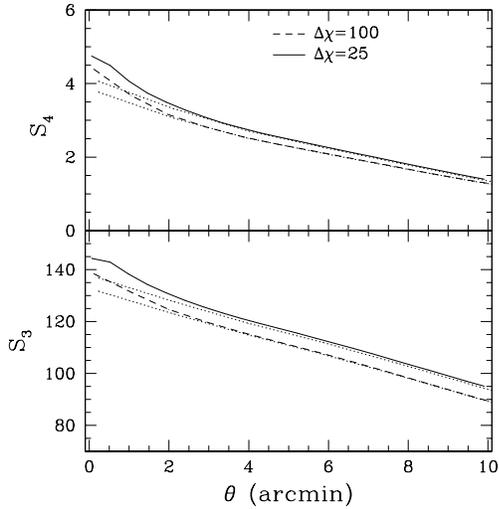}}
\end{center}
\caption{(bottom) The averaged skewness $S_3$ and (top) kurtosis $S_4$ of the
convergence for ten realizations of our high resolution simulation, as a
function of smoothing scale $\theta$.  These are computed for two different
values of inter-plane spacing $\Delta\chi$ for grids of $4096^2$ (solid,
dashed) and $2048^2$ (dotted).  As in Figure~\ref{fig:s2s3s4}, the kurtosis is in units of $10^4$}
\label{fig:s3s4}
\end{figure}

\begin{figure}
\begin{center}
\resizebox{7cm}{!}{\includegraphics{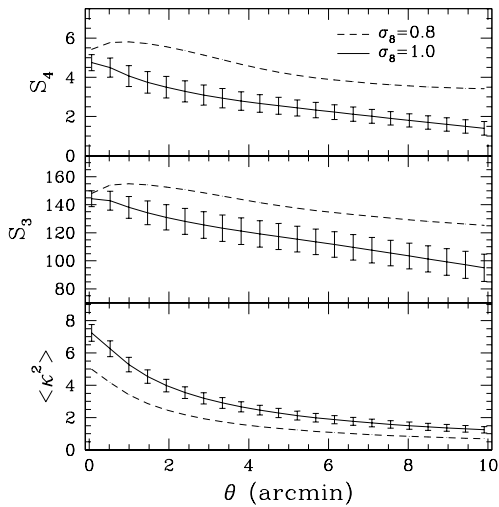}}
\end{center}
\caption{The ten realization average
(bottom) variance, $\langle\kappa^2\rangle$,
(middle) skewness, $S_3$, and
(top) kurtosis, $S_4$, of the convergence for our two high resolution
simulations as a function of smoothing scale $\theta$.
For display purposes, $S_4$ is in units of $10^4$, and
$\langle\kappa^2\rangle$ is in units of $10^{-4}$.}
\label{fig:s2s3s4}
\end{figure}

The maps in Fig.~\ref{fig:kmap} are noticeably non-Gaussian, with identifiable
objects accounting for large positive values of the convergence.  The 2-point
statistics, such as the power spectra, therefore cannot contain all of the
information in the map.  Further cosmological information is contained in
the higher order moments, however the number of degrees of freedom rapidly
becomes large making it difficult to present all of the information.
For this reason we shall present only the `collapsed' N-point functions, and
since the non-Gaussianity is most clearly manifest in real (rather than
Fourier) space we shall present only real-space results.
In Figure~\ref{fig:s3s4} we show the skewness and kurtosis of the convergence
on an angular scale $\theta$.  These are obtained by smoothing the $\kappa$
maps using the IDL command SMOOTH, which performs a real space boxcar average
using square apertures with an integer number of pixels on a side.
We then calculate the skewness
\begin{eqnarray}
  S_3(\theta) = {\langle \kappa^3_{\theta} \rangle \over
                 \langle \kappa^2_{\theta} \rangle^2}
\end{eqnarray}
and the kurtosis
\begin{eqnarray}
  S_4(\theta) =
           {\langle\kappa^4_\theta\rangle - 3\langle\kappa^2_\theta\rangle^2
      \over \langle\kappa^2_\theta\rangle^3}
\end{eqnarray}
by averaging over the pixels in a given field and then averaging the results
for ten realizations.  Individual realizations vary substantially, suggesting
error bars of $\sim 5\%$ for $S_3$ and $\sim 10\%$ for $S_4$, in line with
the results of White \& Hu~(\cite{hhm38}).
Figure \ref{fig:s2s3s4} shows the dependence on scale of the lower order
moments for two cosmological models which differ only in the clustering
strength, $\sigma_8$.

An accurate measurement of the higher order moments, $S_3$ and $S_4$, from
lensing maps may provide important constraints on the cosmological parameters
$\Omega_{\rm m}$ and $\sigma_8$
(Bernardeau, van Waerbeke, \& Mellier~\cite{hhm5}; Jain \& 
Seljak~\cite{hhm18}).
In perturbation theory, the variance of the convergence exhibits a strong
dependence on both quantities, while the skewness and kurtosis are
independent of the latter.
Although the independence of $S_3$ and $S_4$ from $\sigma_8$ is not expected
to hold on scales of interest where the lensing signal can be measured,
a joint measurement of these quantities should still be useful for breaking
the degeneracy between $\Omega_{\rm m}$ and $\sigma_8$ {\it if\/} the actual
dependence can be understood.
This has motivated efforts to extend skewness and kurtosis predictions to
smaller angular scales.

We compare some of these calculations with a variety of numerical results
to assess the degree of convergence between groups and methods.
Table~\ref{tab:s3s4} presents $S_3$ and $S_4$ reported by several groups,
both from simulations and by analytic means.  All are for cosmological models
with $\Omega_{\rm m}=0.3=1-\Omega_\Lambda$ and $h = 0.7$.
The moments from simulations are from JSW, HM (Hamana \& 
Mellier~\cite{hhm12}),
WH (White \& Hu~\cite{hhm38}), H2 (Hamana~\cite{hhm13}), and this paper (VW), 
while the last two rows are analytic predictions.
The first of these, v.W (van Waerbeke~\cite{hhm35}), is for a semi-analytic
fit to N-body simulations which tries to interpolate between the
Hyper-Extended Perturbation Theory (HEPT) prediction
(Scoccimarro \& Frieman~\cite{hhm31}; Hui~\cite{hhm17})
and the perturbation theory result.
The second, TJ (Takada \& Jain~\cite{hhm33}), uses the Halo model approach.

We note that the skewness in the highly non-linear regime is predicted by HEPT
to vary as $\sigma_8^{-0.4}$.  Lowest order perturbation theory predicts that
$S_3$ and $S_4$ would be independent of $\sigma_8$.
If we adjust all of the results to $\sigma_8=0.9$ the disagreements are
lessened (for example using HEPT $S_3(1')$ would scale to 144 for VWa and
to 148 for VWb).
Keeping in mind that the simulation results have uncertainties of 10\% or so,
the agreement between groups is quite good, though more work needs to be done
if we are to obtain precision cosmology from the higher order moments.

\begin{table}
\begin{center}
\begin{tabular}{cc}
 $\kappa_{\rm min}$ & \ \ \#/deg     \\    \hline
 0.05        & \ \  485       \\
 0.10        & \ \  164       \\
 0.15        & \ \   69.0     \\
 0.20        & \ \   34.1     \\
 0.25        & \ \   18.2     \\
 0.30        & \ \   10.3     \\
 0.40        & \ \   3.65     \\
 0.50        & \ \   1.52     \\
\end{tabular}
\end{center}
\caption{The average number of peaks in convergence (with $\kappa >
\kappa_{\rm min}$) per $1^{\circ} \times 1^{\circ}$ field.}
\label{tab:peaks}
\end{table}

Finally it is interesting to look at the extrema of the lensing maps.
We located the convergence peaks for our maps by selecting all pixels with
$\kappa$ greater a selected threshold convergence $\kappa_{\rm min}$, and
then directly comparing the convergence of nearby pixels within a radius
of 0.3 arcminutes.  The largest value was then selected as the peak's center.
Table~\ref{tab:peaks} shows the average number of peaks in a $1^{\circ}
\times 1^{\circ}$ field for selected values of $\kappa_{\rm min}$.
The average profile of these peaks is roughly consistent with that of a
projected NFW profile (Bartelmann~\cite{hhm3}; Navarro, Frenk, \& 
White~\cite{hhm24}).

\section{Numerical issues} \label{sec:numerical}

The numerical convergence of the algorithm described in the previous section is
limited by the size and resolution of the N-body simulation and by the finite
grids used to compute the distortion and deflection of the light rays.  In the
following section, we make use of the power spectrum as a metric in order to
test these limits under various conditions, and to evaluate the basic
ray-tracing implementation against the less approximate versions.   We then
comment on the numerical issues involving the skewness and kurtosis.

\begin{figure}
\begin{center}
\resizebox{7cm}{!}{\includegraphics{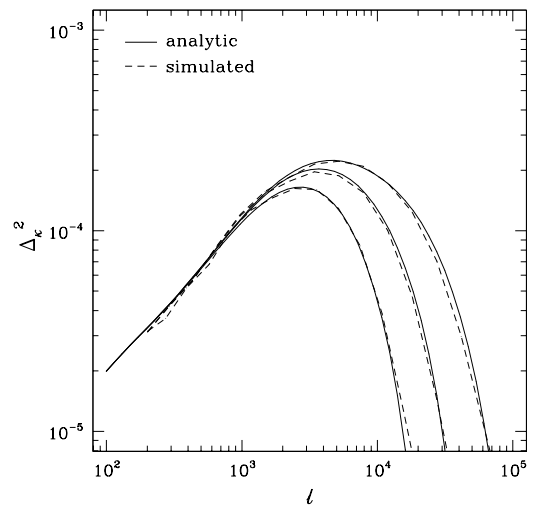}}
\end{center}
\caption{As in the top panel of Figure~\ref{fig:nspectra}, except the
convergence spectra from our high resolution simulation are now compared with
those computed using Eq.~(\ref{eqn:filter}).}
\label{fig:analspectra300}
\end{figure}

\begin{figure}
\begin{center}
\resizebox{7cm}{!}{\includegraphics{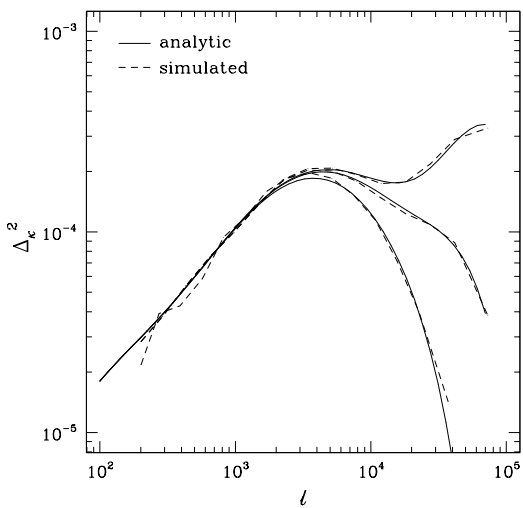}}
\end{center}
\caption{As in Figure~\ref{fig:analspectra300}, except for our low resolution
N-body simulation, which uses $128^3$ particles and has a box side length of
$128 h^{-1}$Mpc.  The spectra are for grids of size $1024^2$, $2048^2$, and
$4096^2$.}
\label{fig:analspectra128}
\end{figure}

\subsection{N-body Simulations}

We use N-body simulations to calculate the evolution of the dark matter
which governs the formation of large-scale structure.  We use 3 simulations
each of a $\Lambda$CDM model.
The first simulation evolves $512^3$ particles in a $300\,h^{-1}$Mpc box,
using the TreePM-SPH code (see the appendix of White~\cite{hhm39}) but without
any SPH particles.  The cosmological model is chosen to provide a
reasonable fit to a wide range of observations with $\Omega_{\rm m}=0.3$,
$\Omega_\Lambda=0.7$, $H_0 = 100\,h{\rm km}\,{\rm s}^{-1}{\rm Mpc}^{-1}$
with $h = 0.7$, $\Omega_B h^2 = 0.02$, $n = 1$ and $\sigma_8 = 1$
(corresponding to $\delta_H = 5.3 \times 10^{-5}$).
The simulation was started at $z = 60$ and evolved to the present with
the full phase space distribution dumped every $100\,h^{-1}$Mpc from
$z\simeq 2$ to $z = 0$.  The gravitational force softening was of a
spline form, with a ``Plummer-equivalent'' softening length of
$20\,h^{-1}$kpc comoving, and the particle mass is
$1.7\times 10^{10}\,h^{-1} M_{\odot}$.
The 3-d mass power spectrum of this simulation is is given for selected
redshifts in Figure \ref{fig:3dspectra}.

We use this simulation as our `fiducial' model unless otherwise specified.
To test the dependence on the amplitude of the power spectrum we also used
a second simulation which is identical to the first except that the
amplitude of clustering is reduced: $\sigma_8=0.8$.
We also make use of a smaller, lower resolution simulation for certain
of the numerical tests described below.  This final simulation is similar
to the larger ones except that it uses fewer ($128^3$) particles and a
smaller ($128\,h^{-1}$Mpc) box and it dumps the output more frequently
(every $25\,h^{-1}$Mpc).  The softening length in this simulation is
$36\,h^{-1}$Mpc.

\subsection{Large scale resolution}

Our ability to resolve large scale features is limited by the field of view of
our ray tracing, and therefore by the size of the box of the N-body simulation
and the distance to the furthest lens plane.  At $z=1$, this limit is 7.5
degrees for our larger simulation and 3.2 degrees for our smaller one.
However, most of our ray tracing runs were made at a field of view of 2
degrees.  Only a few modes are available to contribute to the power spectrum on
large scales for any given realization, leading to substantial variation for
different realizations.  We therefore generate multiple realizations of ray
tracing for an N-body simulation and average them to obtain estimates on large
scales.  For each realization, we project mass randomly along one of three
different axes, and then select the origin of the projected plane at random.
Depending on the distance from the observer, only a fraction of this projected
plane is used in any given ray tracing realization.  For example, a 2 degree
run on our larger simulation uses less than 2\% of a lens plane at the peak of
the lensing kernel.  Multiple nearly independent realizations are therefore
possible using this method, and the results can be averaged to obtain lensing
statistics.

\subsection{Small scale resolution}

The discrete nature of our ray tracing procedure introduces numerical smoothing
when we interpolate mass density values from the N-body simulation onto the
Fourier grid.   The N-body simulation is characterized by both a finite spatial
resolution and by shot noise due to the finite particle number.  To investigate
the effect of this limitation on the angular resolution of the power spectrum,
we model the finite resolution effect with Gaussian filters and thereby modify
Eq.~(\ref{eqn:massspectrum}) so that $\Delta_{\rm m}^2 (k)$ is replaced with
%
%
\begin{equation}
\Delta_{e}^2 (k) \equiv
\left( \Delta_{\rm m}^2 (k) e^{-\sigma_n^2 k^2} +
{k^3 \over 2 \pi^2 \bar n} \right ) e^{-\sigma_{g}^2 k^2}
\label{eqn:filter}
\end{equation}
where $\Delta_{e}^2 (k)$ is an effective 3-d mass power spectrum, $k^3 / 2
\pi^2 \bar n$ is the analytic expression for the shot noise, $\bar n$ is the
mean particle number density, and $\sigma_n$ and $\sigma_g$ are characteristic
resolution limits of the N-body simulation and the Fourier grid, respectively.
We did not find it necessary to introduce a term for shot-noise on each
individual lens plane.

We generate spectra using Eq.~(\ref{eqn:filter}), normalizing the height to
equal that of the simulation at $\ell = 300$.
As shown in Fig.~\ref{fig:analspectra300}, the spectra generated in this
manner using $\sigma_g=0.54 L_{\rm box}/N_{\rm grid}$
(corresponding to a full width half max resolution of 1.3 grid cells)
and $\sigma_n$ set equal to zero are a good fit with those from our high
resolution simulation.
This suggests that for this simulation the power spectrum resolution is
limited by the grid and not by the N-body resolution.  It is also consistent
with our expectation that the Fourier grid's resolution is roughly equal to
its spacing.

The limits imposed by the N-body simulation can be seen in the spectra of our
low resolution simulation (Fig.~\ref{fig:analspectra128}), which are well
approximated by those generated using Eq.~(\ref{eqn:filter}), with
$\sigma_g=0.54 L_{\rm box}/N_{\rm grid}$ as before, and $\sigma_n=50 h^{-1} \rm
kpc$, or 5\% of the mean inter-particle spacing $\bar n^{-{1 \over 3}}$.   The
three curves are for different values of N.  The curve for $N = 1024$ is
similar to those of Figure \ref{fig:analspectra300}, however the roll over is
now caused by both the grid resolution and the N-body resolution.  The curves
at $N = 2048$ and $N = 4096$ show substantially more power at high angular
resolution.  This is because the noise term, which is larger than the signal
from this N-body simulation for $\ell \simeq 10^4$ and above, is no longer
suppressed by the grid resolution.

The value for $\sigma_n$ is independent of grid spacing and is instead related
to the resolution of the simulation.  However, it is not equal to the ``Plummer
equivalent'' softening length, which is $36 h^{-1}$kpc.  To check this, we
computed the 3-d mass power spectrum $\Delta^2_{\rm m}(k)$ using the rebinning
technique of Smith et al.~(\cite{hhm32}) for three simulations with 
identical initial
conditions but varying particle number.  We find that $\sigma_n = 0.05 \ \bar
n^{-{1 \over 3}}$ does roughly reproduce the roll off in power between the
models, though it clearly doesn't get the whole form right because the models
have more power at high-$k$ than this filter suggests.

\begin{figure}
\begin{center}
\resizebox{7cm}{!}{\includegraphics{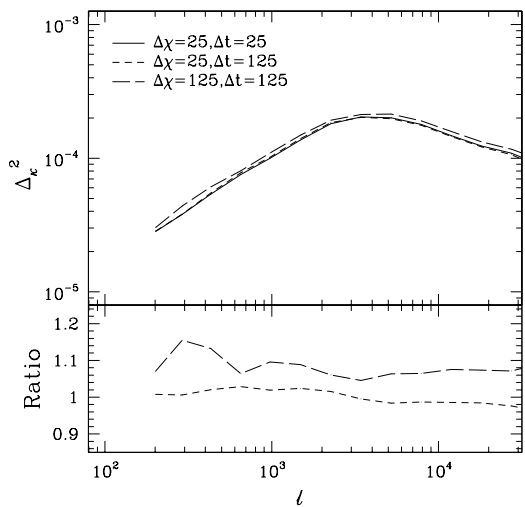}}
\end{center}
\caption{(top) The convergence power spectrum (averaged for 10 realizations) of
our low resolution simulation using different values of comoving distance
$\Delta\chi$ between lens planes and time intervals $\Delta t$ between
particle dumps of the N-body simulation. The units in both cases are
$h^{-1}$Mpc.
(bottom)  The ratio of the two spectra with $\Delta\chi = 125\,h^{-1}$Mpc
with respect to the one with $\Delta \chi = \Delta t = 25\,h^{-1}$Mpc.}
\label{fig:dxdtspectra}
\end{figure}

We make use of our low resolution simulation to investigate the convergence of
the algorithm with respect to the number of lens planes used and to determine
the effect of decreasing the redshift/time interval between particle dumps of
the N-body simulation.  Figure~\ref{fig:dxdtspectra} shows the power spectrum,
averaged over 10 realizations, for different values of inter-plane spacing
($\Delta\chi$) and redshift/time interval ($\Delta t$).
A change in $\Delta\chi$ from $125 h^{-1}$Mpc to $25h^{-1}$Mpc produced a
change in the power spectrum roughly at the $10\%$ level for this simulation
(for our high resolution simulation this value was $\sim 5\%$), however,
additional decreases in $\Delta\chi$ did not significantly improve the
resolution.
For a similar change in $\Delta t$, the spectra are within $1\%$ of each
other.  This is not surprising, since structures evolve on a scale roughly
equal to the Hubble time, which is much longer than the time it takes for
light to travel $100\,h^{-1}$Mpc.

We also wish to understand the numerical resolution of the weak lensing
algorithm as it relates to the higher order moments.  Figure \ref{fig:s3s4}
shows the effect on $S_3$ and $S_4$ of changing the Fourier grid resolution and
the inter-plane spacing $\Delta\chi$.  Increasing the resolution of the grid
increases the small angle power in both the skewness and kurtosis, while
decreasing $\Delta\chi$ increases their power over the full range of
resolution.  However, this effect seems to be due entirely to the decrease in
the convergence variance $\langle\kappa^2\rangle$, while
$\langle\kappa^3\rangle$ and $\langle\kappa^4\rangle$ are essentially
unaffected.

\subsection{Discontinuities}

The process of rotation and origin shifting for each box necessarily leads to
discontinuities in the 3-dimensional mass distributions, which may, in
principle, introduce artifacts during ray-tracing runs.  To investigate this
possibility we traced through our large simulation using the ``angle''  method
described above.  We randomly select the origin and orientation of the first
box and then trace the rays through all the stacked boxes without any further
rotations or origin shifts.  The rays are then initialized in a field of view
about a central line of sight that is angled with respect to the boxes to avoid
repeating any structures, and the lens planes are made so that they are
perpendicular this line of sight.  We averaged the power spectra of ten runs
created using this method and find that it is the same as for the standard
``rotate and shift'' method to better than $1\%$.

\subsection{3-dimensional ray tracing}

One approximation inherent in the lens plane method is that since the light
rays are all traveling in different directions they are only roughly
perpendicular to the lens planes.  The weak lensing formalism is based on
Eq.~(\ref{eqn:einstein}), $d\vec{\alpha}=-2\nabla_\bot\phi\ d\chi$, which
requires that we use the gradient perpendicular to the path of the light-rays.

We used the 3-dimensional ray-tracing algorithm described previously to test
the validity of this approximation, and we find that it holds extremely well
for two degree fields.  We used three-dimensional Fourier grids with
$512\times 512\times 32$ points, with the latter number applying to the
direction of the line of sight.
We find that convergence maps made with the two different methods agree
to better than $0.1\%$, and measures such as the power spectra, skewness,
and kurtosis are virtually indistinguishable.

\section{Physical Issues} \label{sec:physical}

\subsection{The Born approximation near peaks in the convergence}

\begin{figure}
\begin{center}
\resizebox{7cm}{!}{\includegraphics{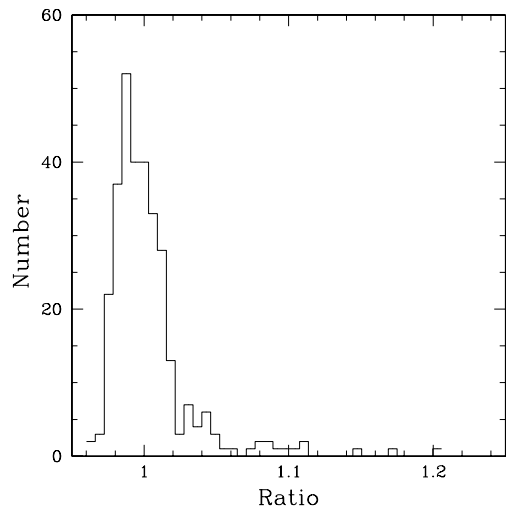}}
\end{center}
\caption{A histogram of the ratio $\kappa/\kappa_B$ of the convergence for
300 large ($\kappa>0.35$) peaks in our maps.  Here, $\kappa$ is the
convergence at the center of a given peak using our full simulation, and
$\kappa_B$ is the convergence of the same peak calculated using the Born
approximation.}
\label{fig:born}
\end{figure}

An approximation that is often useful in weak lensing analysis is the Born
approximation, in which the lensing events for a given light ray are completely
decoupled from one another.  The distortion and deflection contributions for
each event are computed at points along an undeflected path and as if for an
undistorted ray.  These are then summed to produce the total deflection and
distortion.  The Born approximation can be calculated by tracing rays along
straight paths without deflections and approximating ${\bf A}_n$ from
Eq.~(\ref{eqn:discrete}) as
\begin{eqnarray}
{\bf A}_n  = {\bf I} - \sum_{m=1}^{n-1} g(\chi_m , \chi_n) {\bf U}_m
\label{eqn:born}
\end{eqnarray}

The Born approximation has been shown before to be valid for whole convergence
maps using measures such as the power spectrum.
However, a breakdown of this approximation, if one were to occur, would almost
certainly take place where lensing effects are strongest, which could in
principle be an issue of relevance to lensing by clusters of galaxies in
the real universe.
In Fig.~\ref{fig:born} we compare 300 peaks with convergence values of at
least $0.35$ computed using our standard ray-tracing vs.~the same peaks
using the Born approximation.

We find that the Born approximation is nearly always accurate to within 10\%
even for these large peaks, and the single biggest departure was 20\%.  In
addition, the mean of the ratios is very nearly equal to one, so deviations
appear to be unbiased.  We expect that the Born approximation becomes
increasingly accurate as $\kappa$ is decreased, but matching the `Born peaks'
with those of the full ray tracing (including deflections) becomes
increasingly difficult so we have not been able to quantify this.

\subsection{Comparison to a universal magnification PDF}

In Figure~\ref{fig:holzupdf}, we compare the magnification probability
distribution function of a typical run, $P(\mu)$, with the analytical
prediction using the universal probability distribution function $P(\eta)$
(Wang, Holz, \& Munshi~\cite{hhm37}), where
\begin{eqnarray}
  P(\mu) = P(\eta) / 2 | \kappa_{\rm min} |  \qquad .
\end{eqnarray}
The reduced convergence is defined as
$\eta\equiv {1\over 2}(\mu-1)/|\kappa_{\rm min}|+1$,
and  $\kappa_{\rm min}$ is the ``empty beam'' convergence.

The top panel shows a typical PDF from our simulations vs. the prediction from
the model.  The shape of the model PDF agrees with the simulation value to
within $\sim 5\%$ near the PDF peak and $\sim 20\%$ in the high $\mu$ tail.
However, we found that the model PDF was often shifted (in either direction) on
the horizontal axis with respect to that of the simulation.  The magnification
$\mu$ at the PDF peaks was typically within $0.5 \%$ but disagreed by as much
as $1.5 \%$ for extreme cases, such as the one shown here.  This is a
substantial disagreement given the few percent amplitude of the lensing signal.
Even so, the shape of the PDF is still well fit, as can be easily seen if the
model PDF is shifted to lower $\mu$ so that the curves overlap.

\subsection{Multiple lensing events}

\begin{table}
\begin{center}
\begin{tabular}{ccccc}
               &      &$\#$ of Events&      &            \\
$\kappa_m$ \ \ &   1  &    2           &  3        &  4     \\    \hline
  0.01   \ \ \ & 40.8 & 9.78           & 1.61      & 0.19   \\
  0.05   \ \ \ & 3.46 & 0.06           & 0.0005    &$< 10^{-4}$\\
  0.10   \ \ \ & 0.68 & 0.003          &$< 10^{-4}$& 0      \\
  0.15   \ \ \ & 0.21 & 0.0003         & 0         & 0      \\
  0.20   \ \ \ & 0.08 & 0.0001         & 0         & 0      \\
  0.25   \ \ \ & 0.03 & $< 10^{-4}$    & 0         & 0      \\
\end{tabular}
\end{center}
\caption{The number of lensing events above a threshold convergence $\kappa_m$
for 10 runs, given here as a percentage of the roughly 42 million total number
of lines of sight.  Light-rays are unlikely to experience large distortions
more than once, suggesting that peaks in the convergence are almost always the
result of a single lensing event.}
\label{tab:lenslens}
\end{table}

During the ray-tracing runs, we tracked the number of lensing events above
selected threshold magnitudes $\kappa_m$ for each ray.  A lensing event is
defined here as occurring when a ray experiences a shear on a single lens plane
that causes a change in the convergence at the observer in excess of
$\kappa_m$, so that $g(\chi_m,\chi_n){\bf U}_m > \kappa_m$ in
Eq.~(\ref{eqn:discrete}).  In Table~\ref{tab:lenslens}, we present the
results from ten simulations with a total of more than $4 \times 10^7$ lines of
sight.  It is quite uncommon for a given light-ray to experience multiple
lensing events of a large magnitude, which suggests that a given large peak in
the convergence maps is nearly always associated with a single localized
massive object in the N-body simulation.

\section{Conclusions} \label{sec:conclusions}

Weak gravitational lensing has become a powerful tool for observing the
large-scale matter distribution in the universe, with rapid advances in
observational capabilities hinting at even greater things to come.
As observers achieve ever greater precision, it's important that the
theory keep pace.

In this paper, we have described a multi lens-plane algorithm for generating 
maps of weak lensing distortion from structure generated by N-body simulations
paying particular attention to the numerical convergence of the algorithm(s).
While our findings are substantially in line with previous works, we have
noticed that biases can creep in if not enough lensing planes or time dumps
are used.  As concerns the range of validity of the generated maps, we find
that the small-scale spatial resolution is well described by
Eq.~(\ref{eqn:filter}).  With modern, parallel, codes the N-body simulation
is not a limiting factor in lensing studies, suggesting that grids of models
could be run relatively inexpensively.

The basic multi lens-plane algorithm introduces two numerical artifacts to the
calculation of distortion maps (other than finite resolution of the grids used
in computations).  One is the introduction of structure discontinuities at the
box boundaries in the 3-dimensional matter distribution used to make the lens
planes, and the other is that light-rays are not truly perpendicular to the
lens planes.  We have examined the effect of both of these artifacts using techniques
specifically designed to avoid them, by tracing through the boxes at an angle
in the first case, and using a 3-dimensional simulation in the second, and we have
shown that they do not introduce significant changes in the maps.

We examined the applicability of the Born approximation to weak lensing using
measures of the statistical properties of whole maps, such as the power
spectrum and the skewness, and confirm that they are in generally good
agreement with semi-analytic predictions.  We also confirm that the Born
approximation is valid even near peaks in the convergence, where the most
non-linear lensing events in our simulations are likely to occur.

Weak gravitational lensing is a tool that is providing insight into the nature
of our universe at a rate that is expected to increase dramatically in coming
years, and simulations will play a useful role.  We have demonstrated that
simulations of weak lensing based on a simple implementation of the multiple
lens-plane algorithm generate maps of lensing distortion that are a good
approximation of maps created using more computationally intensive methods.
We have also shown that the numerical resolution of these simulations is well
approximated by a simple formula.
This suggests that a relatively modest investment in computation can provide
the highly accurate, simulated maps which are crucial to developing algorithms
and understanding data in this field.

\bigskip
The simulations used here were performed on the IBM-SP2 at the National
Energy Research Scientific Computing Center.
This research was supported by the NSF and NASA.

\end{document}